\documentclass{emulateapj}
\usepackage{amsmath}
\usepackage{graphicx}

\shorttitle{Electron Injection in Non-relativistic Shocks}
\shortauthors{Riquelme \& Spitkovsky}

\begin{document}

\title{Electron Injection by Whistler Waves in Non-relativistic Shocks}

\author{Mario A. Riquelme}
\affil{Astronomy Department, University of California, Berkeley, CA 94720}
\email{marh@astro.berkeley.edu}
\and
\author{Anatoly Spitkovsky}
\affil{Department of Astrophysical Sciences, Princeton University, Princeton, NJ 08544}

\begin{abstract}
Electron acceleration to non-thermal, ultra-relativistic energies ($\sim 10-100$ TeV) is revealed by radio and X-ray observations of shocks in young supernova remnants (SNRs). The diffusive shock acceleration (DSA) mechanism is usually invoked to explain this acceleration, but the way in which electrons are initially energized or `injected' into this acceleration process starting from thermal energies is an unresolved problem. In this paper we study the initial acceleration of electrons in non-relativistic shocks from first principles, using two- and three-dimensional particle-in-cell (PIC) plasma simulations. We systematically explore the space of shock parameters (the Alfv\'{e}nic Mach number, $M_{A}$, the shock velocity, $v_{sh}$, the angle between the upstream magnetic field and the shock normal, $\theta_{Bn}$, and the ion to electron mass ratio, $m_i/m_e$). We find that significant non-thermal acceleration occurs due to the growth of oblique whistler waves in the foot of quasi-perpendicular shocks. This acceleration strongly depends on using fairly large numerical mass ratios, $m_i/m_e$, which may explain why it had not been observed in previous PIC simulations of this problem. The obtained electron energy distributions show power law tails with spectral indices up to $\alpha \sim 3-4$. The maximum energies of the accelerated particles are consistent with the electron Larmor radii being comparable to that of the ions, indicating potential injection into the subsequent DSA process. This injection mechanism, however, requires the shock waves to have fairly low Alf\'{e}nic Mach numbers, $M_A \lesssim 20$, which is consistent with the theoretical conditions for the growth of whistler waves in the shock foot ($M_A \lesssim (m_i/m_e)^{1/2}$). Thus, if the whistler mechanism is the only robust electron injection process at work in SNR shocks, then SNRs that display non-thermal emission must have significantly amplified upstream magnetic fields. Such field amplification is likely achieved by the escaping cosmic rays, so electron and proton acceleration in SNR shocks must be interconnected. 
\end{abstract}

\keywords{acceleration of particles - shock waves - cosmic rays - plasmas}

\section{Introduction}
\label{sec:intro}
Non-thermal electron acceleration is believed to be a universal feature of non-relativistic collisionless shocks both in space plasma and astrophysical environments. In the Earth's bow shock, for example, electrons accelerated up to few tens of keV are usually observed in quasi-perpendicular regions, i.e, where the angle between the upstream magnetic field and the shock normal is $\gtrsim 45^{\circ}$ \citep{OkaEtAl06, GoslingEtAl89}. Also, radio and X-ray observations of supernova remnants (SNRs) show synchrotron radiation produced by relativistic, non-thermal electrons accelerated in their forward shocks \citep[e.g.][]{KoyamaEtAl95, BambaEtAl03, BambaEtAl05}. 

Despite its ubiquity, the actual mechanism for shock acceleration of electrons is still a mystery. The diffusive shock acceleration (DSA) mechanism is the most accepted theory for particle acceleration in shocks \citep{AxfordEtAl77, Krymsky77, Bell78, BlandfordEtAl78}. This theory assumes that particles are scattered by MHD turbulence both in the upstream and downstream regions of the shock. Under these conditions, particles move diffusively in the shock vicinity, gaining energy through many crossings of the shock. However, in order to be able to cross the shock many times, the particles need to have Larmor radii that are larger than the shock width, which is controlled by the typical ion Larmor radius. This is the still unresolved ``injection problem" of the DSA theory, which is particularly stringent for electrons due to their small Larmor radii. 

Several methods have been applied to studying electron acceleration in non-relativistic shocks. One approach, typically used in the context of low Alfv\'{e}nic Mach number ($M_A$) shocks, uses test particle electrons moving in electromagnetic fields pre-determined from hybrid simulations (where ions are modeled as kinetic particles and electrons as a massless fluid). Recently, \cite{GuoEtAl10} showed that electrons can be energized due to repeated shock crossings as they move along pre-existing, large-scale magnetic fluctuations in perpendicular shocks. Also, \cite{Burgess06} used the same method to show that shock ``ripples," which are fluctuations on the surface of quasi-perpendicular shocks, can provide the scattering necessary for efficient electron acceleration. Although in principle these mechanisms may contribute significant electron energization in non-relativistic shocks, they still require a population of electrons that are somehow injected upstream with energies significantly above thermal. Understanding the origin of these particles requires self-consistent kinetic calculations based on particle-in-cell (PIC) simulations. 

PIC simulations have been used by several authors to study the electron acceleration problem in non-relativistic shocks \citep[e.g.,][]{AmanoEtAl07, AmanoEtAl09, UmedaEtAl09}. For instance, using two-dimensional PIC simulations, \cite{AmanoEtAl09} showed that efficient electron acceleration (with spectral index $\alpha=$ 2-2.5) can happen in a perpendicular, $M_A = 14$ shock due to ``shock surfing" of electrons on electrostatic waves from Buneman instability excited at the leading edge of the shock foot. These simulations were done for relatively low mass ratios ($m_i/m_e=25$) and for a particular geometry of the magnetic field (strictly out of the plane of the simulation). The dependence of this mechanism on different shock parameters needs to be clarified, and in fact, in this work we find that this mechanism is not very efficient at realistic mass ratios. 
Also, using two-dimensional PIC simulations \cite{UmedaEtAl09} showed that, for $M_A = 5$ shocks, pre-acceleration in Buneman waves can be complemented by further energization due to scattering at the shock ripples, supporting the picture laid out by \cite{Burgess06}. The energy spectrum of the electrons in this case, however, does not correspond to a  power law tail, but is rather described by two Maxwellian distributions at different temperatures.

One of the main difficulties of PIC studies is that using realistic physical parameters is computationally expensive, so unrealistic approximations have to be made. For instance, simultaneously capturing the dynamics of electrons and ions requires resolving time scales as short as the inverse of the plasma frequency of electrons, $\omega_{p,e}^{-1}$, and as long as the inverse of the cyclotron frequency of ions, $\omega_{c,i}^{-1}$ (with $\omega_{p,j}=(4\pi n_j e_j^2/m_j)^{1/2}$ and $\omega_{c,j}=e_jB/m_jc$, where $c$ is the speed of light, $B$ is the magnitude of the magnetic field, and $n_j$, $m_j$, and $e_j$ are the density, mass, and electric charge of the $j$ species). Given that $\omega_{p,e}/\omega_{c,i} = (m_i/m_e)^{1/2}/(v_A/c)$, where $v_A$ $\equiv B/(4\pi n_i m_i c^2)^{1/2}$ corresponds to the Alfv\'{e}n velocity of the plasma, the computing time in PIC simulations is usually reduced by using artificial mass ratios $m_i/m_e \ll 1836$ and/or relatively large values of $v_A/c$. Also, these studies are usually made using one- and two-dimensional simulations. Although much of the relevant physics can still be revealed using these approximations, a complete understanding of the role played by the chosen parameters is crucial before any extrapolation to realistic setups is made. Thus, in this work we present a systematic exploration of the space of shock parameters, paying special attention to the role of artificial mass ratios $m_i/m_e$ on the possible electron energization.

The effect of using a reduced ion to electron mass ratio $m_i/m_e$ has already been highlighted by previous PIC studies of quasi-perpendicular shock structure \citep{SchoelerEtAl03, SchoelerEtAl04, HellingerEtAl07}. These works have shown that excessively small $m_i/m_e$ may suppress the appearance of oblique whistler waves in the foot of low $M_A$ shocks, which would lead to periodic shock reformation.

In terms of electron energization, the works of \cite{SchoelerEtAl03, SchoelerEtAl04} also show that whistler waves can lead to significant electron heating in the foot of the shocks, but the investigation of their possible role in non-thermal electron injection remains to be realized\footnote{We note that \citet{Levinson92, Levinson94} discussed an analytical theory of electron injection due to whistler waves excited by the returning electrons. These works predicted that electrons should be accelerated in shocks with large Mach numbers, $M_{A}\gtrsim 43$. However, both our simulations and the recent work by \citet{KatoEtAl10} show neither whistler excitation nor significant electron acceleration in such shocks, raising questions about the consistency of the assumptions that went into the theory.}.

In this work we study the injection of non-thermal electrons in quasi-perpendicular shocks, using two- and three-dimensional PIC simulations. We systematically test different regimes for the shock velocity, $v_{sh}$, the plasma magnetization (quantified in $M_A$), the ion to electron mass ratio $m_i/m_e$, and the angle between the shock normal and the ambient magnetic field, $\theta_{Bn}$. We show that the presence of oblique whistler waves in the shock foot does lead to significant electron acceleration with spectral index $\alpha \approx 3-4$. This acceleration appears to prefer small values of the $M_A/(m_i/m_e)^{1/2}$ ratio, and requires $\theta_{Bn} \ne 90^{\circ}$. These two conditions likely explain why this effect was not seen by previous PIC studies of this problem, where strictly perpendicular shocks with fairly low mass ratios were used. For the realistic value $m_i/m_e=1836$, this acceleration would require low Mach numbers, $M_A \lesssim 20$, in order to explain the electron injection fraction inferred from broadband observations of SNRs. Thus, if this mechanism happens to be the only possible solution for electron injection into the DSA process in SNR shocks, the $M_A \lesssim 20$ condition would imply strong amplification of the upstream magnetic field of these shocks.

This paper is organized as follows. In \S \ref{sec:setup} we explain the numerical setup of our simulations. In \S\ref{sec:injection}, we describe the mechanism that brings electrons to non-thermal energies, by analyzing the evolution of the accelerated particles in one of our simulations. In \S \ref{sec:parameters}, we explore different shock parameters, and determine the regimes where electrons acceleration occurs. Our final discussion and conclusions are presented in \S \ref{sec:disconclu}.
 
 \section{Simulation Setup}
 \label{sec:setup}
We use the electromagnetic PIC code TRISTAN-MP \citep{Buneman93, Spitkovsky05} in two and three dimensions\footnote{In both two- and three-dimensional simulations the three components of particle velocities and electro-magnetic fields are tracked.}. A shock wave is produced by reflecting a cold high-speed electron-ion plasma off a conducting wall. The shock forms due to the interaction between the incoming and reflected beams and propagates away from the wall along the $x-$direction. The incoming plasma carries a uniform magnetic field, $\vec{B}_0$, forming an angle $\theta_{Bn}$ with the shock normal (which coincides with the $x-$axis). The shocked plasma stays at rest with respect to the box, so the simulation is done in the downstream frame. In two dimensions the simulation box consists of a rectangle in the $xy$ plane with periodic boundary conditions in the $y-$direction. 

The incoming plasma is injected through a ``moving injector", which recedes from the wall in the $x-$direction at the speed of light. The simulation box is expanded in the $+ x-$direction as the injector approaches the right boundary. This way the memory and computing time are saved, while following the evolution of all the upstream regions that are causally connected with the shock. Further numerical optimization can be achieved by allowing the moving injector to periodically jump backward (i.e., in the $- x-$direction), resetting the fields to its right \citep[see][]{SironiEtAl09}. Since the shock travels much slower than the speed of light, without a jumping injector (i.e., with the injector only moving at $c$) the upstream region would comprise most of the simulation domain. However, since we expect the electron acceleration to occur on scales close to the shock foot (i.e., at a distance comparable to the typical Larmor radius of ions, $R_{L,i}$), keeping an upstream size of a few shock foot lengths should be enough to capture the relevant acceleration physics. Thus, in our simulations the moving injector jumps backward every 2000 time steps, with the first jump happening after 4000-12000 time steps (depending on the length of the shock foot). The injector jumps backward a distance such that its average velocity is close to the shock speed. This method also helps to reduce particle heating by the numerical cold beam instability, which can happen after a long distance propagation of the cold plasma over the numerical grid.

We ran a series of two- and three-dimensional simulations to explore different shock regimes. The run parameters are summarized in Tables \ref{table:2D}, \ref{table:3D}, and \ref{table:2Dperp}. The numerical value of the speed of light is set to $0.45$ cells/time step in all the runs. In the two-dimensional runs presented in the main part of the paper, $\vec{B}_0$ is always in the plane of the simulation ($xy$ plane). This condition is changed in Appendix \ref{sec:appendix}, where the case of $\vec{B}_0$ quasi-perpendicular to the simulation plane is considered. 

\section{The Acceleration Mechanism}
\label{sec:injection}
In this section we disentangle the process that gives rise to the electron energization, focusing on one of our shock simulations where significant acceleration is observed. We use the two-dimensional run 2D-3, with $v_{sh}/c=0.14$ (as seen from the upstream medium), $v_A/c = 1/50$, $m_i/m_e=400$, $M_A=7$, and $\theta_{Bn}=75^{\circ}$ (the other parameters are specified in Table \ref{table:2D}). First we will describe the basic features of the shock, and then we will focus on the process of electron energization.
\subsection{Shock Structure}
\label{sec:shochstr}
\begin{figure}
\centering
\includegraphics[width=8.5cm]{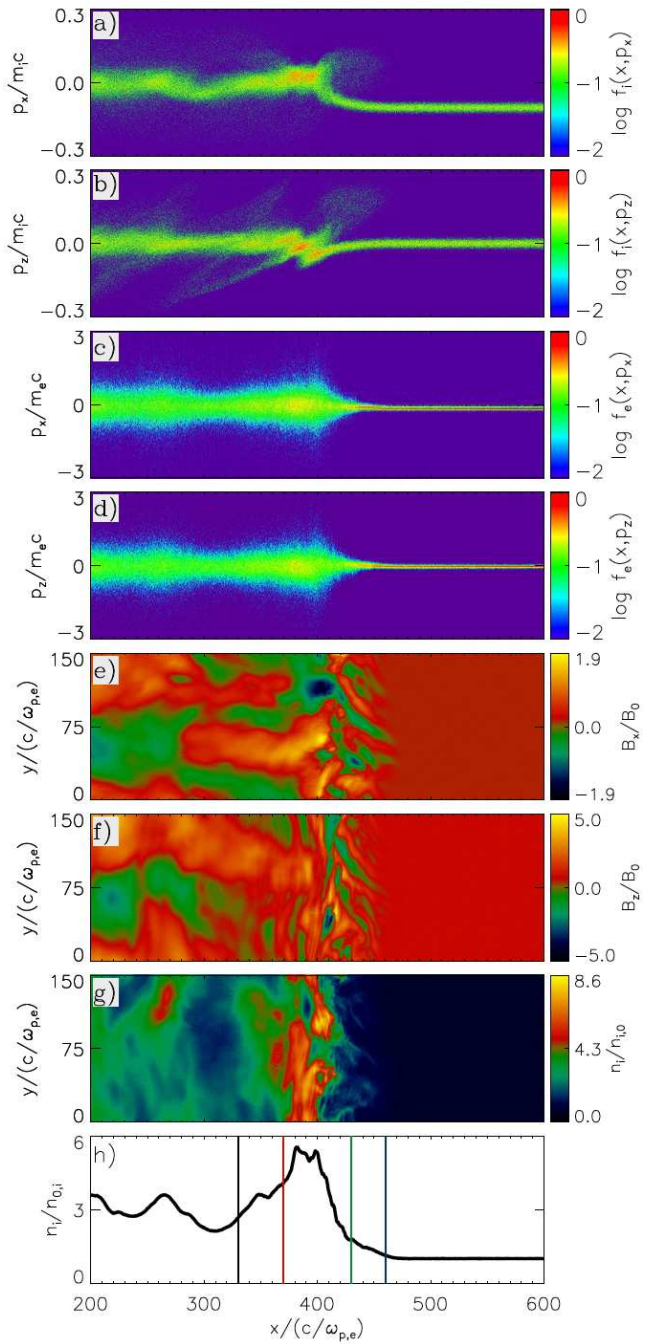}
\caption{The shock structure for run 2D-3 (Table \ref{table:2D}) at $t\omega_{p,e}=10000$ ($t\omega_{c,i}=10$, where $\omega_{c,i}$ is defined in the upstream), whose basic parameters are $v_{sh}=0.14c$, $M_A=7$, $m_i/m_e=400$, and $\theta_{Bn}=75^{\circ}$. The phase space distribution functions of the ions, $f_i$, on the $x-p_{x}$ and $x-p_{z}$ planes are shown in panels $a)$ and $b)$, respectively. The corresponding phase space distribution functions for the electrons, $f_e$, are depicted in panels $c)$ and $d)$, respectively. The distribution functions are normalized by their maximum value. Panels $e)$, $f)$, and $g)$ show the magnetic field components along the $x-$ and $z-$ axes, and the ion density, respectively. Panel $h)$ shows the one-dimensional ion density profile averaged on the $y$ axis.}
\label{fig:phaspace}
\end{figure}
Quasi-perpendicular shocks are characterized by the presence of the so-called shock foot. The foot is defined by the existence of a beam of ions reflected by the shock, whose bulk velocity in the upstream frame is close to $v_{sh}$ and has comparable $x$ and $z$ components. The foot region can be seen in panels $a)$ and $b)$ of Figure \ref{fig:phaspace} (between $x = 410$ and 470 $c/\omega_{p,e}$), which depicts the phase space distribution functions of the ions. This region covers a distance comparable to the ion Larmor radius, $R_{L,i}$ (calculated with the upstream magnetic field $B_0$), and is located right in front of the shock density jump or ``ramp", as shown in panel $h)$ of the same Figure. An important feature of the shock foot is the presence of oblique electromagnetic waves that grow on scales of $\sim 10$ $c/\omega_{p,e}$, forming an angle of $\sim 45^{\circ}$ with the shock normal. These waves can be seen in panels $e)-g)$ of Figure \ref{fig:phaspace} (between $x = 410$ and 470 $c/\omega_{p,e}$), which show the magnetic field along the $x-$direction, $B_x$, the magnetic field along the $z-$direction, $B_z$, and the ion density, $n_i$, respectively. The oblique modes have a right-handed circular polarization, and their phase velocity is comparable to the speed of the shock. All these features allow them to be identified as electron whistler waves propagating obliquely with respect to the background magnetic field.\newline

The whistler waves have been studied in the context of shock structure evolution, using one-, two-, and three-dimensional PIC simulations, as well as hybrid codes. The two-dimensional studies show that whistler waves can play an important role by suppressing the self-reformation of perpendicular and quasi-perpendicular shocks \citep{HellingerEtAl07, LembegeEtAl09, YuanEtAl09}. This effect, however, would be less important in the fully three-dimensional geometry, as shown by a recent PIC study \citep{ShinoharaEtAl11}.

The exact generation mechanism of these whistlers is still subject to debate. One candidate mechanism is the so-called modified two-stream instability (MTSI), driven by the relative cross-field velocity between the electrons and the ions in the foot of the shocks \citep{WuEtAl83, MatsukiyoEtAl03, MatsukiyoEtAl06}. Since in quasi-perpendicular shocks a significant fraction of the ions are reflected into the upstream, the MTSI can be driven by the relative motion between electrons and either the incoming or reflected ions. These two possibilities are usually referred to as MTSI1 and MTSI2, respectively. The analytic dispersion relation calculations show that the MTSI1 and MTSI2 will grow if $\cos(\theta) \gtrsim 4(1-r)M_A/(m_i/m_e)^{1/2}$ and $\cos(\theta) \gtrsim 4 r M_A/(m_i/m_e)^{1/2}$, respectively, where $\theta$ is the angle between the magnetic field and the wave vector of the waves, and $r$ is the fraction of reflected ions, with typical values of $r \sim 0.2$ \citep{MatsukiyoEtAl03}. Independently of which of these two varieties of the MTSI is more dominant, these conditions show that the smaller is the ratio $M_A/(m_i/m_e)^{1/2}$ the larger is the range of $\theta$ where the excitation of whistler waves in the foot of shocks would be possible.\newline 

Another possibility for whistler generation is that these waves can be an intrinsic component of oblique quasi-perpendicular shocks. Indeed, \cite{KrasnoselskikhEtAl02} proposed an analytical model for the structure of these shocks where the shock ramp is treated as a nonlinear whistler wave. This model shows that if $M_A \lesssim M_w\equiv|\cos(\theta_{Bn})|(\sqrt{m_i/m_e})^{1/2}/2$, where $\theta_{Bn}$ is the angle between the shock normal and the upstream magnetic field, $\vec{B}_0$, the shock precursor would contain a stationary whistler wave train. If $M_w$ is exceeded, these whistler waves would become non-linear and would be rather confined to the shock ramp. But if $M_A \gtrsim M_{nw}\equiv|\cos(\theta_{Bn})|(\sqrt{m_i/m_e})^{1/2}/2^{1/2}$ a stationary solution for the shock structure would no longer be possible, which would set the condition for shock reformation. In particular for Run 2D-3 ($\theta \approx 30^{\circ}$ and $\theta_{Bn}=75^{\circ}$) the condition for the whistler generation due to the MTSI (in any of its two varieties) appears to be less stringent than for the model proposed by \cite{KrasnoselskikhEtAl02}. This is in general the case for the simulations presented in this paper. But, even for runs where $\cos(\theta_{Bn}) \gtrsim \cos(\theta)$, the growth of whistler waves in the foot of quasi-perpendicular shocks appears to be favored for small values of the $M_A/(m_i/m_e)^{1/2}$ ratio.\newline

\subsection{Electron Spectrum}
\begin{figure}[!t]
\centering
\includegraphics[width=8cm]{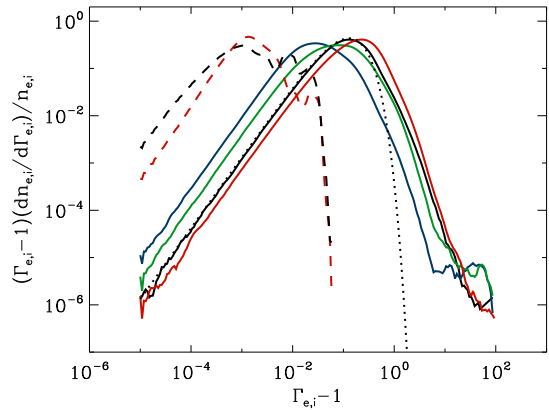}
\caption{The energy spectra at different positions for electrons and ions (solid and dashed lines, respectively) are shown for the shock transition region of run 2D-3 at $t\omega_{c,i}=10$.  The spectra are measured at the $x/c/\omega_{p,e}=330, 370, 430$, and 460 positions, shown in black, red, green, and blue lines, respectively. These positions are also marked by vertical lines in panel $h)$ of Figure \ref{fig:phaspace}, whose colors match the ones used for the corresponding spectra.}
\label{fig:spectpos}
\end{figure}
The phase space distribution functions for electrons is depicted in panels $c)$ and $d)$ of Figure \ref{fig:phaspace}. Although the electrons are mainly heated at the shock, significant electron energization also occurs in the foot region. This is seen from the electron spectra shown in Figure \ref{fig:spectpos}, which are measured at several positions in the shock region. The positions are marked by the vertical lines in Figure \ref{fig:phaspace}h, with colors matching the ones of the corresponding spectra in Figure \ref{fig:spectpos}. We see that the two energy distributions measured in the downstream (red and black lines) show a high-energy power law tail with spectral index $\alpha \simeq 3.6$ (for comparison a Maxwellian distribution corresponding to the downstream electron temperature is shown in black dotted line). The two spectra measured in the foot and the beginning of the ramp (blue and green lines, respectively) do not show a power law tail, but a bump in the high energy part of the distribution. This high energy bump is also observed further upstream, showing that 
some accelerated electrons also outrun the shock as they move along the upstream magnetic field. The maximum Lorentz factor of the electrons  ($\Gamma_e \simeq 50$) is consistent with the electron Larmor radii being comparable to that of the ions, i.e., $\Gamma_e \approx v_{sh}(m_i/m_e)/c$. Also, two downstream ion spectra, measured at the locations of the vertical black and red lines in panel $h)$ of Figure \ref{fig:phaspace}, are depicted by dashed lines. These spectra show an incomplete ion thermalization, and essentially no ion acceleration, with only a small, exponentially decreasing  bump at energies a few times above thermal. Notice that the downstream electron temperature is close to the mean kinetic energy of the ions (for $m_i/m_e=100$), implying  energy equilibration between both species. Finally, the spectra depicted in Figure \ref{fig:spectpos} are fully evolved, in the sense that they do not show significant variations in time. This will be the case in all the simulations shown in this paper after a time of $\sim 6\omega_{c,i}^{-1}$.

\subsection{The Physics of Acceleration}
\label{sec:physaccel}  

\begin{figure*}
\begin{center}
\centering\includegraphics[width=18cm]{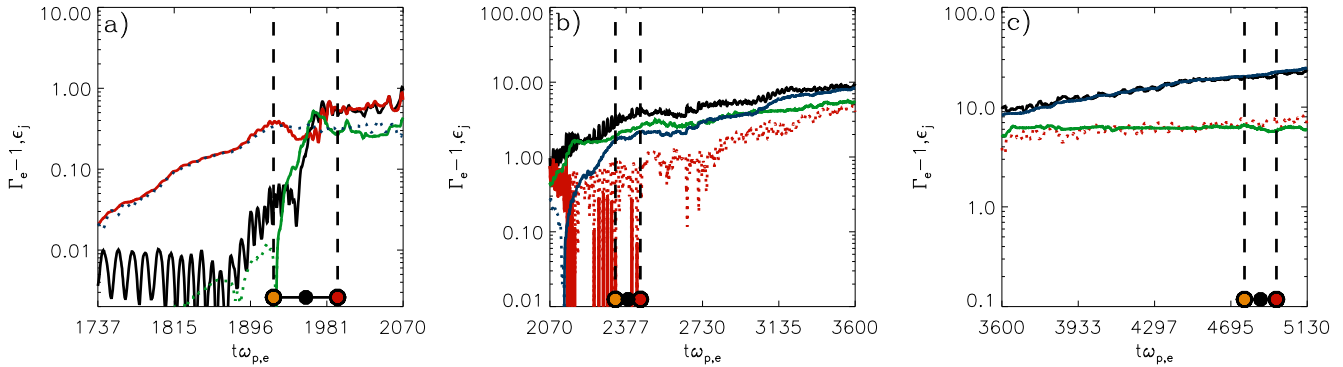}
\caption{The time evolution of the energy of an accelerated electron, $\Gamma_e-1$ (where $\Gamma_e$ is the electron Lorentz factor), is depicted by the solid, black line in panels $a)$, $b)$, and $c)$, which show different time intervals. Besides the electron energy, the accumulated energy gain, $\epsilon_j$, due to the work done by the electric field along the $j$ axis is also plotted (red, green, and blue lines represent $j=x$, $y$, and $z$, respectively). Since these cumulative energies are shown on the logarithmic scale, only their magnitudes are plotted. Thus, in order to keep the sign information, the negative values of  $\epsilon_j < 0$ are plotted using dotted lines. The vertical, dashed lines in panels $a)$, $b)$, and $c)$ mark the intervals corresponding to the electron trajectories tracked in panels $a)$, $b)$, and $c)$ of Figures \ref{fig:elecaccel1} and \ref{fig:elecaccel2}.}
\label{fig:elenevol1}
\end{center}
\end{figure*}

In this section we illustrate the physics of the electron acceleration process by focusing on the evolution of a typical non-thermal electron. In the three panels of Figure \ref{fig:elenevol1} we plot the energy of the electron, represented by the solid, black line. In order to identify the source of the energy gain, we also plot the accumulated energy gain, $\epsilon_j$, due to the work done by the electric field along the $j$ axis (red, green, and blue lines represent $j=x$, $y$, and $z$, respectively). 

\begin{figure}
\centering
\includegraphics[width=8cm]{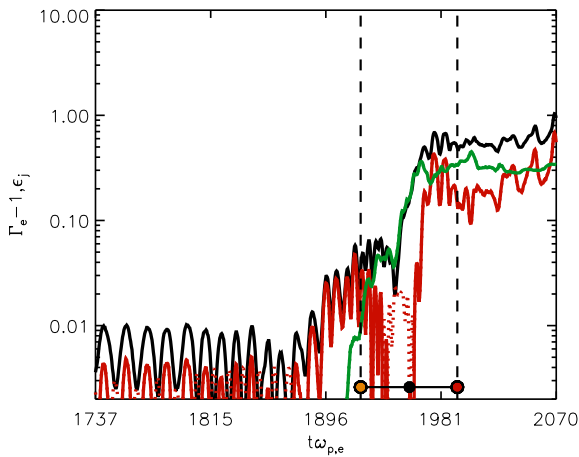}
\caption{The time evolution of the energy $\Gamma_e-1$ (where $\Gamma_e$ is the electron Lorentz factor) of the accelerated electron analyzed in \S \ref{sec:physaccel}, for the same interval shown in panel $a)$ of Figure \ref{fig:elenevol1} is plotted in black solid line. Besides the total energy, the cumulative energy gain due to the work done by the electric field perpendicular and parallel to $\vec{B}$ ($\epsilon_{\perp}$ and $\epsilon_{||}$, respectively) is also plotted in red and green lines, respectively.}
\label{fig:elenevol2}
\end{figure}

\begin{figure}[!t]
\centering
\includegraphics[width=8.5cm]{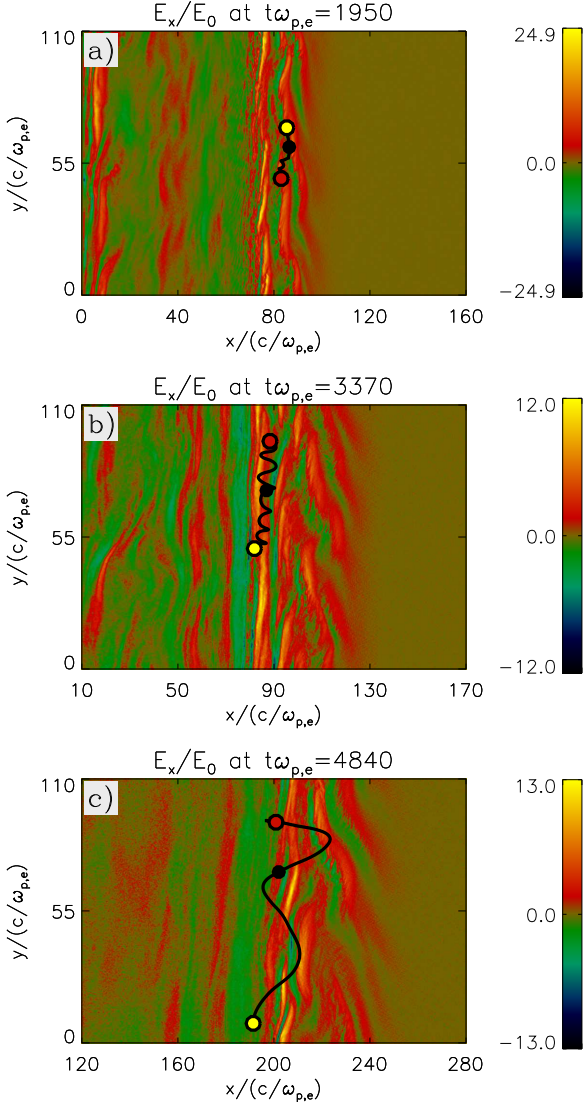}
\caption{The electric field along $\hat{x}$, $E_x$, for the two-dimensional run 2D-3 ($M_A=7$, $v_{sh}=0.14c$, $\theta_{Bn}=75^{\circ}$, and $m_i/m_e=400$). Panels $a)$, $b)$, and $c)$ correspond to times $t\omega_{p,e}=1950$, 3370, and 4840. The field is normalized in terms of $E_0\equiv B_0v_{in}/c$, where $v_{in}$ is the speed at which the upstream plasma is injected, as seen from the downstream medium (for run 2D-3, $v_{in}=0.1c$).  On top of each panel,  the trajectory of an accelerated electron is tracked by the black line, with the yellow and red circles marking the initial and final time of each trajectory. The black circle, marks the time of each field snapshot.}
\label{fig:elecaccel1}
\end{figure}

\begin{figure}[!t]
\centering
\includegraphics[width=8.5cm]{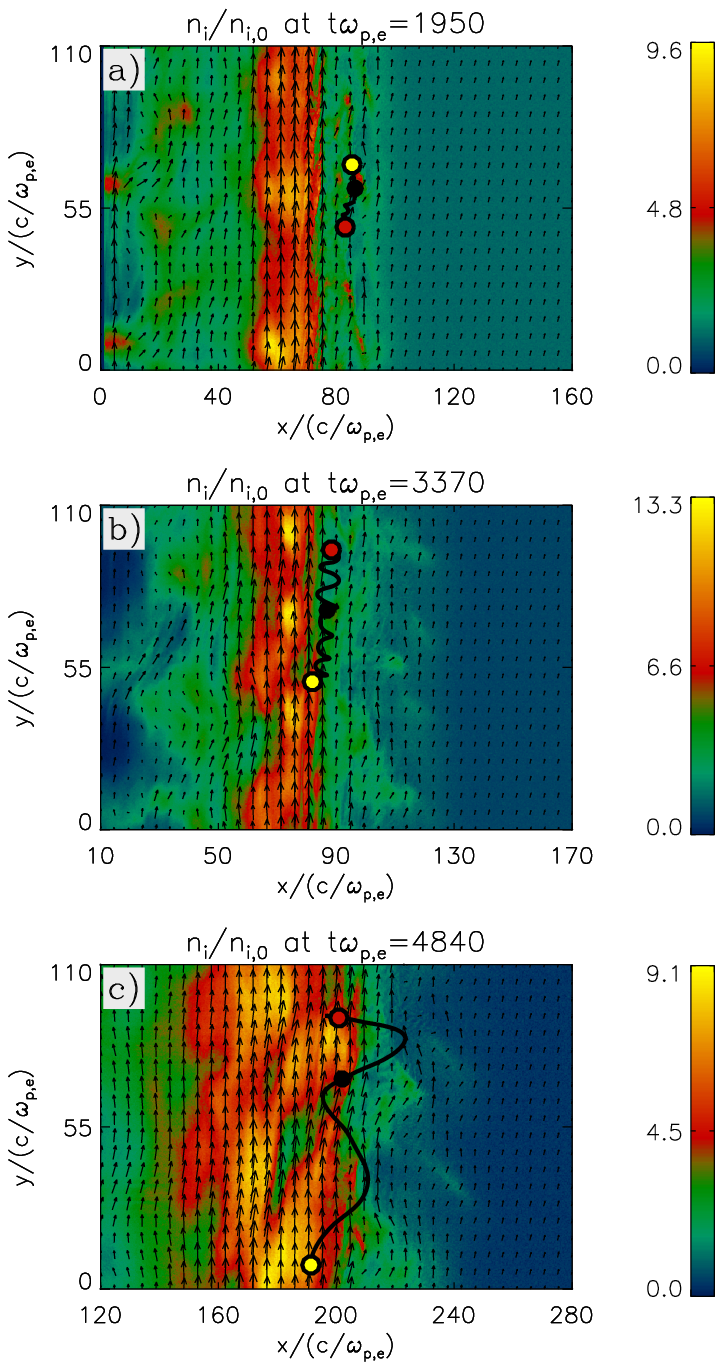}
\caption{Same as in Figure \ref{fig:elecaccel1} but for the ion density, $n_i$. The density is normalized in terms of the upstream ion density, $n_{i,0}$. The arrows represent the orientation of the magnetic field on the $xy$ plane.}
\label{fig:elecaccel2}
\end{figure}
This energy gain is calculated as $\epsilon_j \equiv \int v_{e,j}E_j e/m_e dt$, where $v_{e,j}$ and $E_j$ correspond to the electron velocity and electric field along the $j$ axis, respectively. Since these cumulative energies are shown on the logarithmic scale, we plot their absolute values, and use dotted lines when they correspond to negative quantities. In each panel we also highlight a small time interval, whose starting and final points are marked by yellow and red circles, respectively. The particle trajectories corresponding to each of these intervals are shown on top of Figures \ref{fig:elecaccel1} and \ref{fig:elecaccel2} by a solid, black line. The black circle on top of the highlighted intervals marks the time of each field snapshot. 
The initial increase in the electron energy happens in the first highlighted interval, shown in Figure \ref{fig:elenevol1}$a$,  with the particle location depicted in Figures \ref{fig:elecaccel1}$a$ and \ref{fig:elecaccel2}$a$. We can see that this initial energization happens in the shock foot as the particle moves through the region of amplified whistler waves. The energy gain is initially due to increase in $\epsilon_y$, with $\epsilon_x$ contributing most of the energy gain after $t\omega_{p,e} \approx 1960$. Although before the initial energy growth (the one dominated by $\epsilon_y$) $\epsilon_x$ has already increased significantly, its contribution to the total energy of the electron is almost exactly compensated by a decrease in $\epsilon_z$ (see rising dotted blue and solid red lines between $t\omega_{p,e} =1737$ and 1920 in Figure \ref{fig:elenevol1}$a$). Indeed, the particle motion until this point was dominated by the  $\vec{E}\times \vec{B}$ drift, which does not allow any net work to be done by the electric field. However, as soon as electric field fluctuations on scales comparable to the electron Larmor radius appear, this almost perfect cancellation stops, and a net energy gain becomes possible. Since the $y-$axis is quasi-parallel to the initial field $\vec{B}_0$, the initial energization due to $E_y$ already suggests an important feature of the whistlers waves: their electric field has a non-negligible component along the magnetic field direction. This parallel electric field component can be explained by the obliquity of the waves with respect to the initial magnetic field, which makes the projection of whistler electric field on $\vec{B}_0$ have a magnitude $\sim \sin(\theta)|\vec{E}|$. The energy gain due to the parallel electric field can be verified from Figure \ref{fig:elenevol2}, where we have plotted in red and green the energy gain due to the electric field perpendicular and parallel to $\vec{B}$ ($\epsilon_{\perp}$ and $\epsilon_{||}$, respectively). Indeed, we can see that  $\epsilon_{||}$ is what dominates the initial energy gain of the electron, confirming that the presence of the electric field parallel to the magnetic field is the essential component in the initial electron energization. This feature makes this process fundamentally different from fast Fermi acceleration where the shock acts as a magnetic mirror, energizing the electrons due to the electric field component perpendicular to the magnetic field. Finally, we note that this initial energization is described in the frame where the downstream plasma is at rest (usually called normal incidence frame). The picture, however, should be essentially the same if described from any other frame like, for instance, the de Hoffman-Teller frame. In particular, the initial energization will continue to be dominated by the electric field component parallel to the magnetic field. This is due to the relativistic invariance of $\vec{E} \cdot \vec{B}$, which would make $\vec{E} \cdot \vec{B}/|\vec{B}|$ a nearly invariant quantity under any non-relativistic frame transformation.
  
After the initial energization driven by the whistlers, the acceleration due to $E_z$ (the convective electric field) starts to play a more relevant role, as can be seen from Figure \ref{fig:elenevol1}$b$. We see that in the highlighted interval (marked by the two vertical, dashed lines) the energy increase is driven by a jump both in $\epsilon_z$ and $\epsilon_y$. In Figures \ref{fig:elecaccel1}$b$ and \ref{fig:elecaccel2}$b$ we can see that during these energy increases the electron moves mainly along the $+y-$direction, and stays close to the shock ramp. The relevance of the energy gain due to $E_z$ is more obvious at later times, when the electron Larmor radius has become comparable to a sizable fraction of the foot length. As shown in Figure \ref{fig:elenevol1}$c$, most of the energy is gained due to an increase in $\epsilon_z$. The particle trajectory corresponding to the marked interval (between the two vertical, dashed lines) is plotted in Figures \ref{fig:elecaccel1}$c$ and \ref{fig:elecaccel2}$c$. We can see that, as it moves (mainly along the $+y-$direction), the particle jumps a couple of times from the upstream to the downstream and vice versa, gaining energy in a way similar to shock drift acceleration. \newline 

Apart from dominating the initial electron energization, the energy gain along the magnetic field, $\epsilon_{||}$, also increases the particle's ability to move along the magnetic field lines. This is important because if an electron velocity along $\vec{B}_0$ becomes larger than $v_{sh}/\cos(\theta_{Bn})$, it will be possible for the particle to move along the $+x-$direction faster than the shock. When this happens, the electron will be able to escape upstream, increasing its chances to be further scattered by the whistlers waves. This qualitative picture already shows that having $\theta_{Bn} \ne 90^{\circ}$ is an important component in the acceleration process. This angle dependence of the acceleration is confirmed in the following section where the importance of the shock parameters: $\theta_{Bn}$,  $m_i/m_e$,  $v_{sh}$, and $M_A$ for electron acceleration is studied. Finally, it is important to point out that the evolution of thermal electrons is fundamentally different to the one of the non-thermal particles, which is what we described in this section. Thermal electrons gain some energy due to a random combination of the parallel and perpendicular electric field mainly in the ramp and shock transition region, and then are rapidly incorporated into the downstream medium.

\section{Electron Acceleration Regime}
\label{sec:parameters}

In this section, we seek to determine the physical conditions under which the electron acceleration due to whistler waves happens. To do this, we analyze the acceleration behavior by varying different shock parameters one by one.

\subsection{Angle Dependence}
\label{sec:angledep}
\begin{figure}[!t]
\centering
\includegraphics[width=8.5cm]{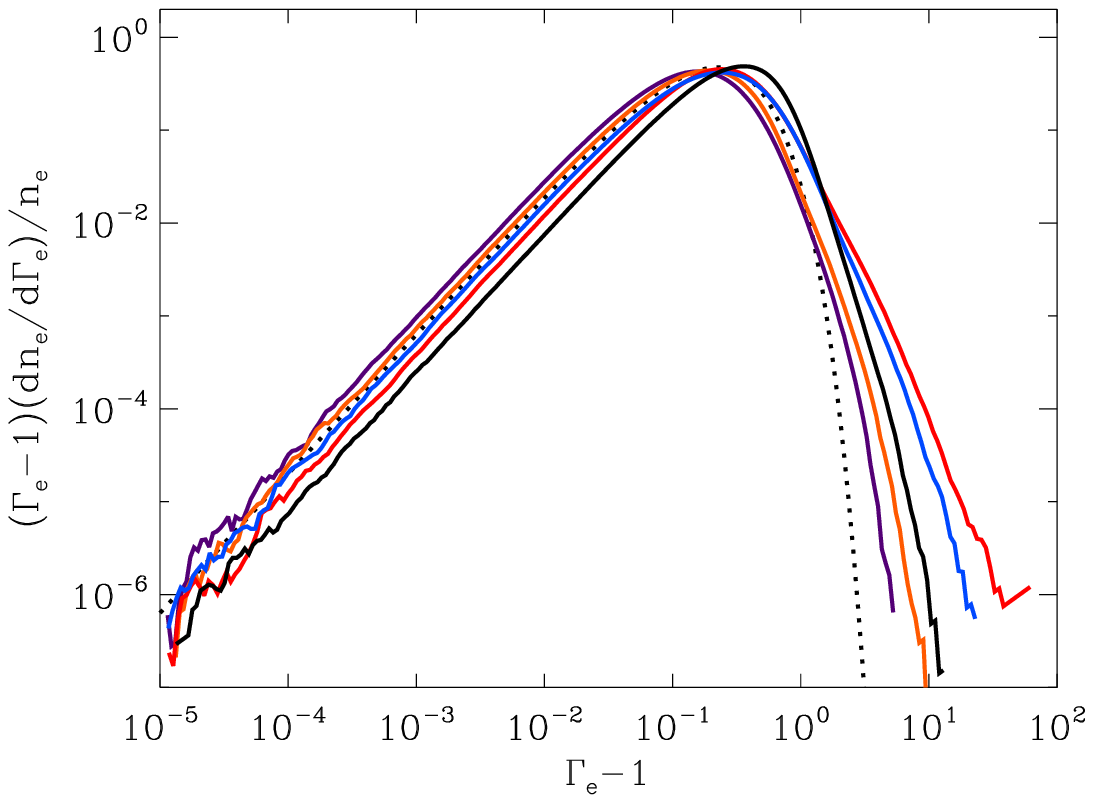}
\caption{The downstream spectrum of the electrons at $t\omega_{p,e}=10000$ ($t\omega_{c,i}=10$) is plotted for simulations like run 2D-3 ($M_A=7$, $v_{sh}=0.14c$, $\theta_{Bn}=75^{\circ}$, and $m_i/m_e=400$), but using different angle $\theta_{Bn}$. The purple, orange, red, blue, and black lines represent the cases $\theta_{Bn}=90$ (run 2D-1), 80 (run 2D-2), 70 (run 2D-4), 60 (run 2D-5), and 45 (run 2D-6). For comparison a Maxwellian distribution is shown in black dashed line.}
\label{fig:anglespect}
\end{figure} 

We illustrate the importance of the angle $\theta_{Bn}$ for electron acceleration using Figure \ref{fig:anglespect}. The downstream spectrum of the electrons at $t\omega_{p,e}=10000$ ($t\omega_{c,i}=10$) is plotted for simulations that only differ in their angle $\theta_{Bn}$. The rest of the parameters are the same as for simulation 2D-3 in Table \ref{table:2D}. We can see that the hardest spectrum is reached for $\theta_{Bn}=70^{\circ}$ (red line), with $\alpha \approx 3.6$ while the softest spectrum is obtained for the $\theta_{Bn}=90^{\circ}$  case (purple line), which corresponds to a Maxwellian distribution with a small increase in the number of particles at energies a few times above thermal. This result confirms the picture suggested in \S \ref{sec:physaccel}, in which the electron energization strongly depends on the ability of these particles to propagate in front of the shock as they slide along the magnetic field lines. This condition implies that the most accelerated particles must move preferentially along the $+y-$axis. This is in fact what we noticed when tracking the trajectory of the accelerated electron analyzed in \S \ref{sec:physaccel}, which represents the typical behavior of the most energetic particles in our runs. For $\theta_{Bn} < 70^{\circ}$ the softer spectra can be explained by the less efficient confinement of the particles to the shock vicinity. A quasi-perpendicular configuration keeps the electrons close to the shock, not allowing them to easily escape upstream. From that perspective, it is reasonable to think that the most efficient acceleration will happen when $\cos(\theta_{Bn})$ is just large enough to allow the electrons to move at $v_{sh}$ in the $+x-$ direction, with decreasing efficiency for larger values of $\cos(\theta_{Bn})$. 

\subsection{Mass Ratio Dependence}
\label{sec:massdep}
\begin{figure}[!t]
\centering
\includegraphics[width=8cm]{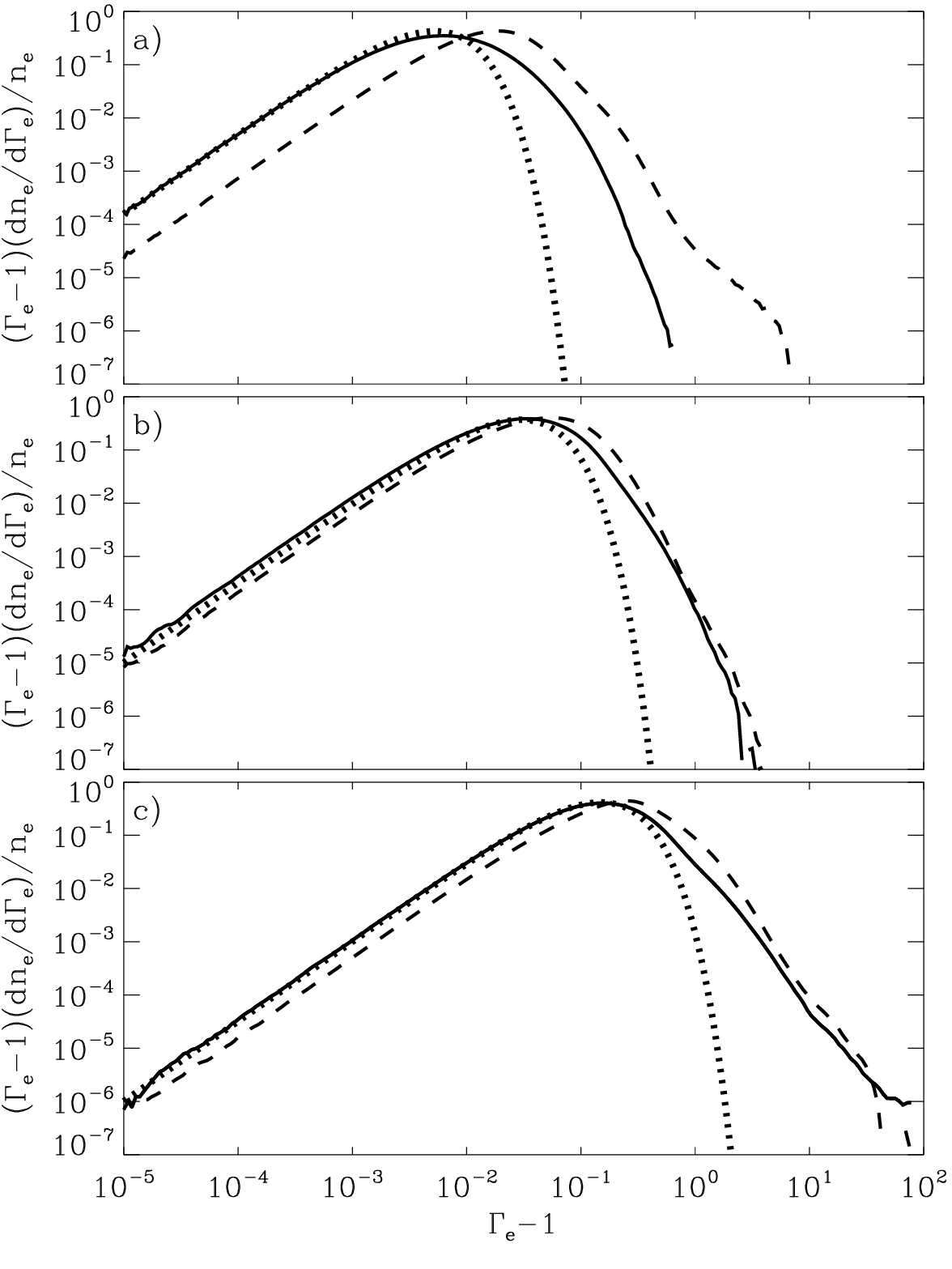}
\caption{The downstream electron spectra for simulations with $M_A=7$, $\theta_{Bn}=75^{\circ}$, and $v_{sh}=0.14c$, but different ion to electron mass ratios, $m_i/m_e=25$, 100, and 400, are depicted in plots $a)$, $b)$, and $c)$, respectively. The solid lines represent two-dimensional runs, while the dashed lines correspond to their three-dimensional counterparts. The parameters of the two-dimensional simulations are described in Table \ref{table:2D} (runs 2D-7, 2D-8, and 2D-3 for $m_i/m_e=25$, 100, and 400, respectively) and the ones of the three-dimensional runs are specified in Table \ref{table:3D} (runs 3D-1, 3D-2, and 3D-3 for $m_i/m_e=25$, 100, and 400, respectively). A convergence between the two- and three-dimensional results can be seen as $m_i/m_e$ increases to more realistic values, implying a decrease in the contribution of the ``shock surfing" mechanism to the acceleration of electrons.}
\label{fig:masspect.b0.1.Ma5}
\end{figure}

\begin{figure}[!t]
\centering
\includegraphics[width=8.5cm]{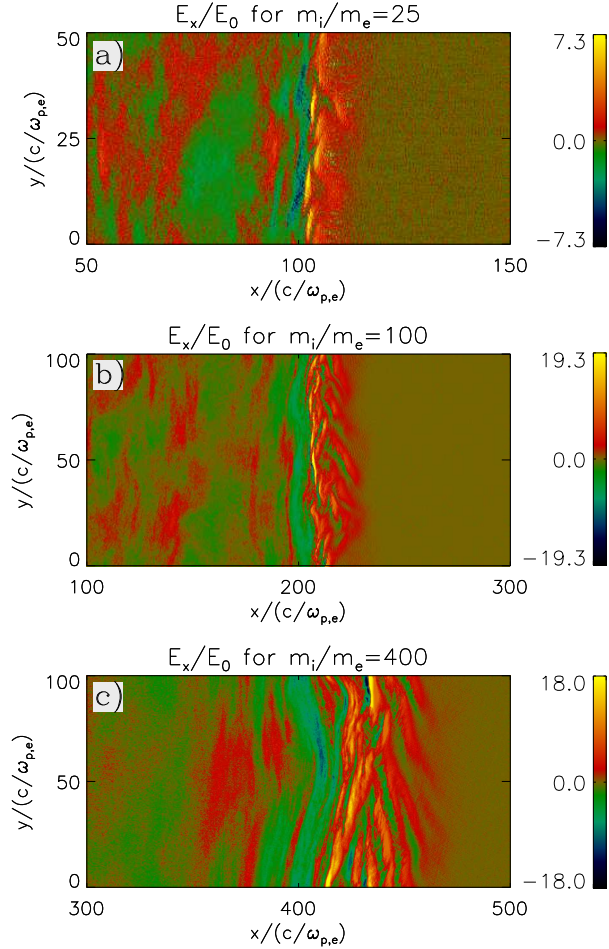}
\caption{The electric field along $\hat{x}$, $E_x$, in the shock transition region of simulations 2D-7, 2D-8, and 2D-3, which only differ in their $m_i/m_e$ parameter given by $m_i/m_e=$25, 100, and 400, respectively. The field is shown at $t\omega_{c,i}=10$ ($t\omega_{p,e}=2500$, 5000, and 10000, respectively), and is normalized in terms of $E_0$, which corresponds to the convective electric field given by $B_0v_{in}/c$ (where $v_i=0.1c$ is the injection velocity of the upstream plasma as seen from the downstream frame).}
\label{fig:masscompare}
\end{figure}

We now explore the ion to electron mass ratio $m_i/m_e$ dependence of the electron acceleration. To do this, we use three runs in two dimensions with the same parameters $M_A=7$, $\theta_{Bn}=75^{\circ}$, and $v_{sh}=0.14c$, but with $m_i/m_e=25$, 100 and 400 (run 2D-7, 2D-8, and 2D-3 of Table \ref{table:2D}, respectively). Their downstream spectra are shown by the solid lines in Figures \ref{fig:masspect.b0.1.Ma5}$a$-\ref{fig:masspect.b0.1.Ma5}$c$. We can clearly see the role played by $m_i/m_e$ in the electron acceleration. While for the mass ratios $m_i/m_e=25$ and 100 cases, the spectral index $\alpha \approx 5.4$ and $\approx 4.7$, respectively, the $m_i/m_e=400$ case shows the decrease of the spectral index to $\alpha\approx 3.6$ \footnote{Since the typical time scale for quasi-perpendicular shock evolution is given by the ion cyclotron period, $\omega_{c,i}^{-1}$, the comparison between simulations of different mass ratio is performed at a fixed time $t\omega_{c,i}=10$. We use this criterion because, for all the mass ratios in our study, $t\omega_{c,i}=10$ corresponds to the typical time at which the downstream spectrum becomes homogeneous, with no substantial variations as a function of position.}. In order to show the correlation between this acceleration and the growth of the whistler modes, in Figure \ref{fig:masscompare} we have plotted the electric field along $\hat{x}$, $E_x$, in the shock transition region of these three simulations (panels $a)$, $b)$, and $c)$ for $m_i/m_e =$25, 100, and 400, respectively). These plots show how the shock foot becomes dominated by large amplitude whistler waves as the mass ratio increases from $m_i/m_e=25$ to 400. Also this Figure confirms the length scale of the whistlers waves ($\sim 10c/\omega_{p,e}$) determined in \S \ref{sec:injection}, which is consistent with previous dispersion relation calculations \citep{WuEtAl83}. 
In this study of mass ratio dependence we also want to include the possibility of electron energization by ``shock surfing" of electrons, due to electrostatic waves produced by the Buneman instability in the foot of the shocks \citep{AmanoEtAl09}. These waves appear in the leading edge of the shock foot on scales of $\sim c/\omega_{p,e}$, and are due to the relative velocity between the upstream electrons and the ions reflected by the shock \citep{Buneman58}. As electrons get scattered by these waves, they gain energy due to the work performed by the fluctuating electric field, combined with the convective field of the upstream plasma. Since the wave vector of the fastest growing Buneman mode, $\vec{k}_{bun}$, is parallel to the ion beam, the growth of the waves and their effect on the electron acceleration are better resolved when the ion motion is parallel to the plane of the simulation ($xy$ plane). This is achieved if the magnetic field, $\vec{B}_0$, is quasi-perpendicular to the $xy$ plane, implying that the Buneman acceleration will be suppressed in our two-dimensional runs with the magnetic field in the simulation plane \citep{AmanoEtAl09}. We used this behavior to distinguish the contribution of the ``shock surfing" mechanism to electron acceleration by comparing the results of the two-dimensional simulations with analogous three-dimensional runs, where the $\vec{k}_{bun}$ is resolved by adding a third dimension of a few $c/\omega_{p,e}$. The corresponding downstream spectra of the three-dimensional runs with $m_i/m_e=25$, 100, and 400 (runs 3D-1, 3D-2, and 3D-3 of Table \ref{table:3D}, respectively) are depicted using dashed lines in Figures \ref{fig:masspect.b0.1.Ma5}$a$, \ref{fig:masspect.b0.1.Ma5}$b$, and \ref{fig:masspect.b0.1.Ma5}$c$. We can see significant differences between the two- and three-dimensional spectra for $m_i/m_e=25$, implying an important contribution to acceleration due to the shock surfing mechanism. Indeed, when analyzing the trajectories of the energetic particles of the three-dimensional run with $m_i/m_e=25$, we find that most of their energy is gained as the electrons get scattered by Buneman waves in the leading edge of the foot. The difference between two- and three- dimensional runs, however, is substantially reduced for the case $m_i/m_e=100$, and almost disappears for  $m_i/m_e=400$, showing that the relative importance of Buneman acceleration, compared to the energization due to whistler waves, decreases for more realistic mass ratios. Further details on this mass ratio dependence of the shock surfing mechanism are given in Appendix \ref{sec:appendix}, where the acceleration of electrons is studied for two-dimensional runs with $\vec{B}_0$ quasi-perpendicular to the plane of the simulation. The results presented in Appendix \ref{sec:appendix} also demonstrate that the acceleration due to whistler waves disappears if the direction along $\vec{B}_0$ is not resolved by the simulation, which makes the two-dimensional configuration used in the main part of this paper (i.e., with $\vec{B}_0$ in plane), the most suitable for the study the electron injection due to whistlers. Also, the convergence between the two- and three-dimensional runs depicted in Figure \ref{fig:masspect.b0.1.Ma5}$c$ shows that whistler acceleration is well modeled by our two-dimensional simulations, with no appreciable changes due to the additional third dimension of a few $c/\omega_{p,e}$. This result, however, does not rule out significant three dimensional effects when a scale comparable to a few whistler scales (i.e., a few 10 $c/\omega_{p,e}$) is resolved along the third dimension. We leave this possibility as a subject of future investigation.

\subsection{Shock Velocity Dependence}
\label{sec:veldep}
\begin{figure}[!t]
\centering
\includegraphics[width=8cm]{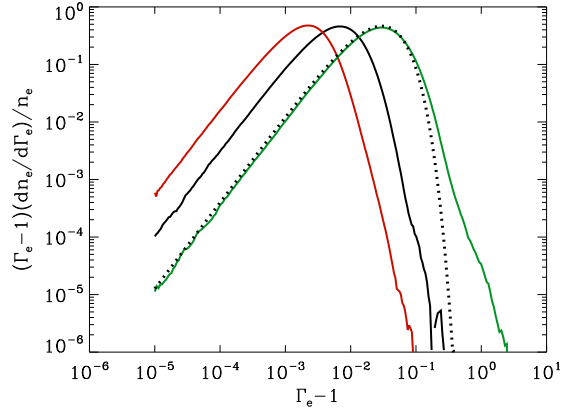}
\caption{The downstream electron spectra for the 2D-18, 2D-19, and 2D-20 simulations, with $v_{sh}=0.042$, $M_A=7$, $\theta_{Bn}=75^{\circ}$, and $m_i/m_e=$25, 100, and 400 (depicted by red, black, and green lines, respectively). A dependence of the spectral index $\alpha$ on $m_i/m_e$ similar to the one found for the $v_{sh}=0.14$ case is observed.}
\label{fig:masspect.b0.03.Ma5}
\end{figure}

It is interesting to see how this electron acceleration may change if a different shock velocity is used. We test the shock velocity dependence by running  simulations with the same parameters as for the two-dimensional simulations shown in Figure \ref{fig:masspect.b0.1.Ma5} but using $v_{sh}=0.042 c$, which corresponds to $v_{in}=0.03 c$ (simulations 2D-18, 2D-19, and 2D-20 in Table \ref{table:2D}). Notice that in order to keep the same $M_A=7$ the magnetic field also has to be reduced by a factor of 3.3. The resultant downstream electron spectra are shown in Figure \ref{fig:masspect.b0.03.Ma5} for $m_i/m_e=25$, 100, and 400 (red, black, and green lines). These spectra essentially reproduce the ones corresponding to two-dimensional runs with $v_{sh}=0.14c$, shown in Figure \ref{fig:masspect.b0.1.Ma5}. The same spectral index $\alpha$ dependence on $m_i/m_e$ is obtained, with a decrease in $\alpha$ when passing from $m_i/m_e=100$ to 400. The overall normalization of the power law tails, however, seems to drop by a factor of $\sim 3$ with respect to the $v_{sh} = 0.14c$ cases. This result suggests that the mechanism for electron acceleration due to growth of whistler waves has a rather weak dependence on the shock velocity, as long as $M_A$ is kept constant. It is important to notice, however, that these measured spectral indices are for the same inclination angles that produce the lowest $\alpha$ in the case of $v_{sh}=0.14c$ ($\theta_{Bn}=75^o$). Thus, in this case we are assuming that the angle $\theta_{Bn}$ at which $\alpha$ is minimized does not depend on the shock velocity. This assumption needs to be confirmed by further exploration of the angle dependence of acceleration for different shock velocities. A thorough study aimed to clarify this point will be presented elsewhere. 

\subsection{Alfv\'{e}nic Mach Number Dependence}
\label{sec:machdep}
\begin{figure*}[t!]
\centering
\includegraphics[width=18cm]{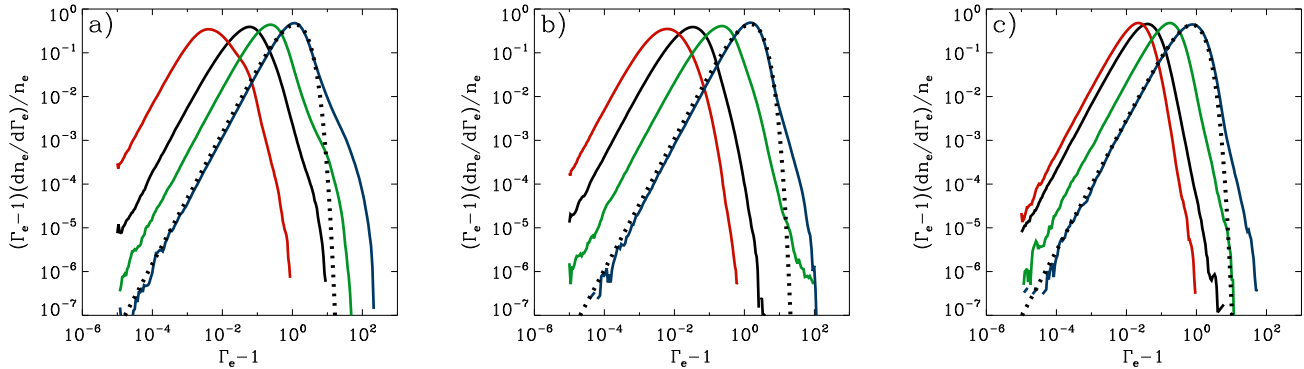}
\caption{The downstream electron spectra are shown for simulations with $M_A=3.5$, 7, and 14, in panels $a)$, $b)$, and $c)$, respectively. The variations in $M_A$ are obtained by only changing the magnitude of the magnetic field, so in all the cases the shock velocity has the value $v_{sh}=0.14c$. The realistic mass ratio $m_i/m_e=1600$ was also included, so the red, black, green, and blue lines represent cases with $m_i/m_e=25$, 100, 400, and 1600, respectively. We can see from these spectra the overall trend to have smaller spectral index $\alpha$ either when the Mach number is reduced or when the mass ratio is increased. The parameters of each of these simulations are compiled in Table \ref{table:2D} (where for $m_i/m_e=25$, 100, 400, and 1600 the $M_A=3.5$ simulations are called respectively 2D-10, 2D-11, 2D-12, and 2D-13, the $M_A=7$ simulations are called 2D-7, 2D-8, 2D-3, and 2D-9, and the $M_A=14$ simulations are called 2D-14, 2D-15, 2D-16, and 2D-17).}
\label{fig:threeMamasspect.b0.1}
\end{figure*}
\begin{figure}[t!]
\centering
\includegraphics[width=8cm]{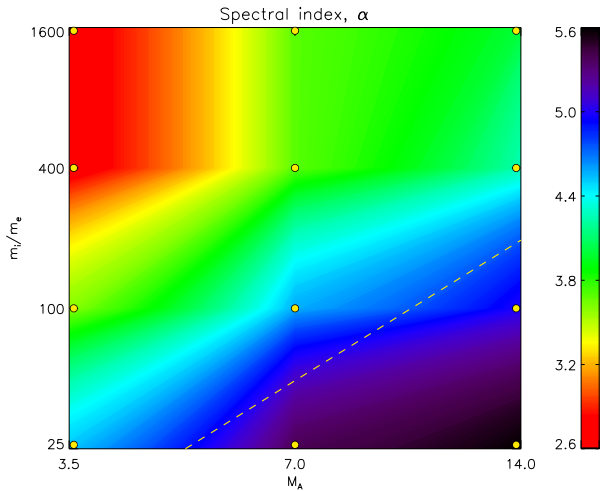}
\caption{A compilation of the spectral indices for the 12 simulations depicted in Figure \ref{fig:threeMamasspect.b0.1} shown as a function $M_A$ and $m_i/m_e$. The yellow dots correspond to the actual grid point locations, while the rest of the diagram is colored using linear interpolation of the measured values of $\alpha$. The yellow dashed line corresponds to $M_A=(m_i/m_e)^{1/2}$. We see that regions of hard and soft spectra tend to be separated by this line, which marks the theoretical estimate for the limit between growth and suppression of whistler waves \citep{WuEtAl83, MatsukiyoEtAl03, KrasnoselskikhEtAl02}.} 
\label{fig:diagram}
\end{figure}
We test the effect of varying $M_A$ by running two-dimensional simulations that are analogous to the ones shown in Figure \ref{fig:masspect.b0.1.Ma5}, but also using the Mach numbers $M_A=3.5$ and $M_A=14$. We do this by changing the magnitude of the magnetic field while keeping the same shock velocity $v_{sh}=0.14c$. We use the mass ratios $m_i/m_e=25$, 100, 400, and 1600, in order to simultaneously take into account the mass ratio dependence for each of the studied Mach numbers. The parameters of the simulations are summarized in Table \ref{table:2D} (where for $m_i/m_e = 25$, 100, 400, and 1600 the $M_A = 3.5$ simulations are called respectively 2D-10, 2D-11, 2D-12, and 2D-13, the $M_A = 7$
simulations are called 2D-7, 2D-8, 2D-3, and 2D-9, and the $M_A = 14$ simulations are called 2D-14, 2D-15, 2D-16, and 2D-17). 

The corresponding downstream electron spectra for $M_A=3.5$, 7, and 14 are shown in panel a), b), and c) of Figure \ref{fig:threeMamasspect.b0.1} (with the red, black, green,  and blue lines showing the spectra for $m_i/m_e=25$, 100, 400, and 1600, respectively). We see that, apart from the hardening of the spectra due to the increase in the mass ratio, the spectral index $\alpha$ also drops as the Mach number goes from 14 to 3.5. This tendency is consistent with our finding that the electron acceleration is driven by the growth of whistler waves in the foot of quasi-perpendicular shocks. Indeed, our analysis presented in \S \ref{sec:shochstr} showed that the growth of whistler waves requires $M_A/(m_i/m_e)^{1/2} \lesssim 1$ \citep{WuEtAl83, MatsukiyoEtAl03, KrasnoselskikhEtAl02}. In Figure \ref{fig:threeMamasspect.b0.1} we see that when this condition is not satisfied (for instance, when $M_A=7$ and $m_i/m_e=25$, shown by the red line in plot $b$) we obtain a rather soft spectrum, with $\alpha \sim 6$. The acceleration efficiencies, in terms of the fraction of non-thermal particles, follow a similar trend, with the number of non-thermal particles being larger for smaller values of $M_A/(m_i/m_e)^{1/2}$. The typical fraction of non-thermal particles in our runs ranges between $\sim 2$ to $10\%$, with corresponding energy fractions of $\sim 10$ to $30\%$\footnote{These percentages are estimated by considering the electrons with energies in the range where the power-law component of the spectrum is larger than the thermal component.}.

For each Alfv\'{e}nic Mach number, we explored the angle dependence of the electron acceleration by using several values for $\theta_{Bn}$. Thus, the spectra presented in Figure \ref{fig:threeMamasspect.b0.1} correspond to the hardest spectra for each $M_A$, whose angles are: $\theta_{Bn}=75^o$ for $M_A=3.5$, $\theta_{Bn}=75^o$ for $M_A=7$, and $\theta_{Bn}=60$ for $M_A=14$. However, these angles were determined using a single mass ratio $m_i/m_e=400$. Thus our results assume that the angle of the hardest spectral index depends only weakly on the used mass ratio. This assumption needs to be confirmed by further exploration of the angle dependence of this acceleration mechanism for different values of $m_i/m_e$. Despite this uncertainty, our results show, in general, a hardening of the electron spectra as the $M_A/(m_i/m_e)^{1/2}$ ratio decreases. This can also be seen from Figure \ref{fig:diagram}, which shows a compilation of the spectral indices for the cases depicted in Figure \ref{fig:threeMamasspect.b0.1} as a function of the mass ratio and Mach number. The yellow dashed line, which corresponds to $M_A=(m_i/m_e)^{1/2}$, clearly separates the regions of small $\alpha$ from regions of rather soft spectra. In the cases with realistic mass ratio ($m_i/m_e=1600$) the spectral index goes from $\alpha=2.7$ to 4 as $M_A$ changes from 3.5 to 14. We will discuss the consequences that this result has for electron acceleration in SNR shocks below.      

Finally, we emphasize that this acceleration mechanism does not show a significant dependence on the chosen value of $\beta_e$. This is verified by directly comparing run 2D-3 and 2D-21, which only differ in their $\beta_e$ ($\beta_e$=0.5 and 0.05, respectively) and show essentially no difference in their final electron spectra. 
\begin{deluxetable*}{ccccccccc}
\tabletypesize{\scriptsize}
\tablecaption{Parameters for the two-dimensional simulations}
\tablehead{ \colhead{Run} &  \colhead{$c/\omega_{p,e}$}&  \colhead{$L_y/(c/\omega_{p,e})$}  &  \colhead{$\beta_e=\beta_i$}&\colhead{$v_{sh}/c$} &
 \colhead{$M_A$}  &  \colhead{$\theta_{Bn}$} &  \colhead{$m_i/m_e$} &  \colhead{N$_{ppc}$}
}
\startdata
2D-1 &  10&  102&  0.5& 0.14 & 7 & $90^{\circ}$ & 400& 100\\
2D-2 &  10&  102&  0.5& 0.14 & 7 & $80^{\circ}$ & 400& 100\\
2D-3 &  10&  102&  0.5& 0.14 & 7 & $75^{\circ}$ & 400& 100\\
2D-4 &  10&  102&  0.5& 0.14 & 7 & $70^{\circ}$ & 400& 100\\
2D-5 &  10&  102&  0.5& 0.14 & 7 & $60^{\circ}$ & 400& 100\\
2D-6 &  10&  102&  0.5& 0.14 & 7 & $45^{\circ}$ & 400& 100\\
2D-7 &  15&  34&  0.5&0.14 & 7 & $75^{\circ}$ & 25& 100\\
2D-8 &  10&  77&  0.5& 0.14 & 7 & $75^{\circ}$ & 100& 100\\
2D-9 &10&  102&  0.05& 0.14 & 7 & $75^{\circ}$ & 1600& 10\\
2D-10 &10&  102&  0.01& 0.14 & 3.5 & $75^{\circ}$ & 25& 10\\
2D-11 &10&  102&  0.01& 0.14 & 3.5 & $75^{\circ}$ & 100& 10\\
2D-12 &10&  102&  0.01& 0.14 & 3.5 & $75^{\circ}$ & 400& 10\\
2D-13 &10&  102&  0.01& 0.14 & 3.5 & $75^{\circ}$ & 1600& 10\\
2D-14 &10&  102&  0.2& 0.14 & 14 & $60^{\circ}$ & 25& 10\\
2D-15 &10&  102&  0.2& 0.14 & 14 & $60^{\circ}$ & 100& 10\\
2D-16 &10&  102&  0.2& 0.14 & 14 & $60^{\circ}$ & 400& 10\\
2D-17 &10&  102&  0.2& 0.14 & 14 & $60^{\circ}$ & 1600& 10\\
2D-18 &  15&  34& 0.005& 0.042 & 7 & $75^{\circ}$ & 25& 100\\ 
2D-19 &  15&  51& 0.005& 0.042 & 7 & $75^{\circ}$ & 100& 32\\
2D-20 &  10&  77& 0.005&0.042 & 7 & $75^{\circ}$ & 400& 65\\
2D-21 &  10&  102&  0.05& 0.14 & 7 & $75^{\circ}$ & 400& 100
\enddata
\tablecomments{We list the electron skin depth $c/\omega_{p,e}$ in terms of number of grid cells, the transverse size of the simulation box in terms of $c/\omega_{p,e}$, the beta parameter of the different plasma particles $\beta_j$ ($\equiv p_j/B_0^2/8\pi$, where $p_j$ is the pressure of particle ``$j$"), the upstream medium shock velocity, $v_{sh}$, the Alfv\'{e}nic Mach number, $M_A$, the angle between the upstream magnetic field and the shock normal, $\theta_{Bn}$, the ion to electron mass ratio, $m_i/m_e$, and the total number of particles per cell, $N_{ppc}$, in the simulation.}
\label{table:2D}
\end{deluxetable*}

\begin{deluxetable*}{ccccccccc}
\tabletypesize{\scriptsize}
\tablecaption{Parameters for the three-dimensional simulations}
\tablehead{ \colhead{Run} &  \colhead{$c/\omega_{p,e}$}&  \colhead{$L_y\times L_z/(c/\omega_{p,e})^2$}  &  \colhead{$\beta_e=\beta_i$}&\colhead{$v_{sh}/c$} &
 \colhead{$M_A$}  &  \colhead{$\theta_{Bn}$} &  \colhead{$m_i/m_e$} &  \colhead{N$_{ppc}$}
}
\startdata
3D-1 &  5&  $26\times3$& 0.5& 0.14 & 7 & $75^{\circ}$ & 25& 100\\ 
3D-2 &  5&  $77\times3$& 0.5& 0.14 & 7 & $75^{\circ}$ & 100& 25\\
3D-3 &  5&  $51\times3$& 0.5& 0.14 & 7 & $75^{\circ}$ & 400& 50
\enddata
\tablecomments{We list the electron skin depth $c/\omega_{p,e}$ in terms of number of grid cells, the transverse size of the simulation box in terms of $c/\omega_{p,e}$, the beta parameter of the different plasma particles $\beta_j$ ($\equiv p_j/B_0^2/8\pi$, where $p_j$ is the pressure of particle ``$j$"), the upstream medium shock velocity, $v_{sh}$, the Alfv\'{e}nic Mach number, $M_A$, the angle between the upstream magnetic field and the shock normal, $\theta_{Bn}$, the ion to electron mass ratio, $m_i/m_e$, and the total number of particles per cell, $N_{ppc}$, in the simulation.}
\label{table:3D}
\end{deluxetable*}

\section{Summary and Conclusions}
\label{sec:disconclu}
In this paper we studied the problem of electron acceleration in non-relativistic electron-ion shocks, using two- and three-dimensional PIC simulations. By systematically exploring the space of shock parameters, and using realistic ion to electron mass ratios, we identify a new acceleration mechanism, that is able to accelerate electrons starting from fairly low temperatures. Thus, it constitutes a possible candidate to solve the well known electron ``injection problem" of the diffusive shock acceleration (DSA) theory. This mechanism confirms the idea that non-thermal electron acceleration can be caused by electron scattering due to foot waves, which are produced by the electron-ion counter-streaming \citep[see, for example, ][]{CargillEtAl88}. We have described the physics of this mechanism, and identified the physical regime where this acceleration is most efficient. The acceleration occurs preferentially in quasi-perpendicular shocks, and is driven by oblique whistler waves excited in the shock foot.

We found that, for simulations that only differ in their ion to electron mass ratio $m_i/m_e$, the spectral index tends to harden as the $m_i/m_e$ grows from $25$ to $1600$. This trend is confirmed for $M_A=3.5$, 7, and 14, and for shock velocities $v_{sh}=0.14c$ and $0.042c$. On the other hand, simulations that only differ in their Alfv\'{e}nic Mach numbers also show a progressive hardening of their spectra as $M_A$ is reduced from 14 to 3.5. 
This mass ratio/Alfv\'{e}nic Mach number dependence of the acceleration is consistent with theoretical arguments suggesting that the growth of whistler waves in the foot of quasi-perpendicular shocks would be favored when $M_A \lesssim (m_i/m_e)^{1/2}$. Although the physics of this whistler excitation is still a subject of debate, this condition appears to hold independently of whether the whistlers are generated by the MTSI \citep{WuEtAl83, MatsukiyoEtAl03, MatsukiyoEtAl06}, or if they are just explained as an intrinsic component of the structure of quasi-perpendicular shocks \citep{KrasnoselskikhEtAl02}. We also found a strong dependence of the acceleration on the angle between the magnetic field and the shock normal, $\theta_{Bn}$, which needs to satisfy $\theta_{Bn} \ne 90^{\circ}$. On the other hand, we found that the shape of the spectra do not depend significantly on the shock velocity, although a factor of $\sim 3$ decrease in the normalization of the non-thermal part of the distribution was observed when the shock velocity was reduced from $v_{sh}=0.14c$ to $v_{sh}=0.042c$.

This dependence of the acceleration on the shock parameters can be explained in terms of two requirements: $1)$ the shock needs to be able to excite whistler waves in the foot, which translates into the condition $M_A \lesssim (m_i/m_e)^{1/2}$, and $2)$ the accelerated electrons need to stay in the shock foot for a time long enough to increase their chances of being  scattered by the whistler waves. The second requirement explains why the efficiency of the acceleration drops substantially when $\theta_{Bn} = 90^{\circ}$. In that case, energized electrons cannot move along the shock normal, $\hat{n}$, since they are tied to the magnetic field. Thus, they are rapidly advected into the downstream medium of the shock. This does not happen in an oblique shock, where accelerated particles can develop a significant velocity along $\hat{n}$. Given that the growth of whistler waves does not depend on the shock velocity, $v_{sh}$, the weak dependence of the acceleration on $v_{sh}$ implies that the ability of the accelerated electrons to stay in the shock foot is also independent of $v_{sh}$. We can verify this point as follows. The initial electron energization is driven by the electric field of the whistler waves (in particular, its component parallel to the magnetic field). This electric field has a typical length scale of $\sim 10 c/\omega_{p,e}$ and a maximum amplitude of $\sim 10 E_0$ (where $E_0 \equiv B_0v_{sh}/c$). Thus, the energy gain in each scattering will be given by $\sim 100 \textrm{ } e E_0 \textrm{ } c/\omega_{p,e}$. Therefore, an electron that moves along $\vec{B}_0$ with that energy, will have $x-$velocity larger than the shock velocity if $100 \textrm{ } e E_0 \textrm{ } c/\omega_{p,e} \cos(\theta_{Bn})^2 \gtrsim m_e v_{sh}^2$, which implies $100 \cos(\theta_{Bn})^2 \gtrsim M_A/(m_i/m_e)^{1/2}$. We can see that this condition is independent of the shock velocity, and confirms that $\theta_{Bn}$ and the $M_A/(m_i/m_e)^{1/2}$ ratio are the crucial parameters for the acceleration mechanism.

\begin{figure}[!t]
\centering
\includegraphics[width=8cm]{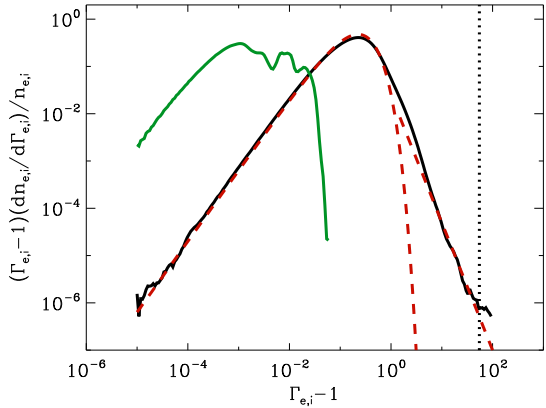}
\caption{The downstream energy spectra for electrons and ions (black and green lines, respectively) are shown for the shock transition region of run 2D-3 ($M_A=7$, $v_{sh}=0.14c$, $\theta_{Bn}=75^{\circ}$, and $m_i/m_e=400$). The thermal and power law tail fits are shown in red dashed lines. The vertical dotted line marks the electron Lorentz factor corresponding to equal electron and ion Larmor radii ($\Gamma_e = (m_i/m_e)(v_{sh}/c)$).}
\label{fig:spectconclu}
\end{figure}

The maximum electron energy that we measure is consistent with the electron Larmor radii $R_{L,e}$ being comparable to that of the ions $R_{L,i}$, as can be verified from Figure \ref{fig:spectconclu}, where we plot the downstream spectrum of electrons for run 2D-3 ($M_A=7$, $v_{sh}=0.14c$, $\theta_{Bn}=75^{\circ}$, and $m_i/m_e=400$). The electron spectrum is shown as black line, and the thermal and power law tail fits are depicted in red dashed lines ($\alpha \approx 3.6$). The vertical dotted line marks the electron Lorentz factor corresponding to $R_{L,e} = R_{L,i}$, which coincides with the maximum energy of the electrons. However, this maximum energy is statistically limited by the number of macroparticles used in the simulations. Thus, higher energy electrons may in principle be possible. In Figure \ref{fig:spectconclu} we also depict the corresponding downstream ion spectrum, which shows incomplete thermalization and the absence of a power law tail. 

We also explored the contribution to acceleration given by the previously proposed ``shock surfing" mechanism, which is driven by the excitation of Buneman waves in the leading edge of the foot \citep{AmanoEtAl09}. We find that its contribution is strongly dependent on the artificial mass ratio $m_i/m_e$, becoming negligible for realistic values of $m_i/m_e$.

From the observational viewpoint, our results are consistent with in-situ electron spectrum measurements in the Earth's bow shock, in which typically $M_A \approx 5-10$. For instance, \cite{GoslingEtAl89} reported on spectrum measurements performed on the International Sun-Earth Explorer (ISEE) 1 and 2 spacecraft, showing that non-thermal electrons are commonly found in the downstream medium of the quasi-perpendicular portions of the shock, but are rarely observed in the quasi-parallel portions. They also found typical spectral indices of $\alpha \approx 3-4$, with the power-law part of the spectrum extending smoothly out of the thermal part of the distribution, which is essentially what we observe in our simulations. Similar results were more recently reported by \cite{OkaEtAl06}, using data from the Geotail spacecraft. We emphasize, however, that the application of our results to the Earth's bow shock measurements is based on simulations with $v_{sh}=0.14c$ and $0.042c$, which are a factor of at least $\sim 100$ larger than the shock velocities typically found in the solar system. This extrapolation is made due to the finding that the crucial parameters for the acceleration due to whistlers are $M_A$ and $\theta_{Bn}$, with the shock velocity dependence being much weaker. This extrapolation remains to be confirmed by shock simulations with orders of magnitude smaller $v_{sh}$ (or $v_A$).

In the context of electron acceleration in SNR shocks, we can confirm the potential importance of this injection mechanism by estimating the Alfv\'{e}nic Mach number necessary to explain the typical fraction of particles injected into the DSA process, $\eta_{inj}$. According to the modeling of broadband observations of SNRs, $\eta_{inj}$ ranges between $\sim 10^{-4}$, for magnetic fields amplified only by shock compression ($B \sim 17 \mu G$), and $\sim 10^{-6}$, for significant magnetic amplification consistent with recent X-ray observations ($B \sim 130 \mu G$) \citep[see, for example,][]{ZirakashviliEtAl10}. Also, in order to be injected, the particles need to have Larmor radii close to the size of the shock transition region, whose characteristic scale corresponds to the Larmor radii of the ions. Thus, we can estimate the maximum slope of the non-thermal part of the spectrum necessary to satisfy these requirements. If we assume that the normalization of the non-thermal spectrum is such that the power law tail dominates for energies very close to the peak of the thermal distribution (which is approximately what we see in our simulations), we get that the necessary spectral index should satisfy
\begin{equation}
\alpha \approx 1 - \log(\eta_{inj})/\log(c/v_{sh}).
\end{equation}
Thus, for a typical shock velocity of $v_{sh} = 3000$ km/s, and $\eta_{inj}=10^{-6}$, we get that $\alpha \approx 4$. From the results summarized in Figure \ref{fig:diagram}, we see that, for realistic mass ratios ($m_i/m_e=1600$), the maximum Mach number that would give $\alpha \lesssim 4$ corresponds to $M_A=14$. This estimate shows that the injection mechanism presented in this paper is a viable solution to the injection of electrons into the DSA mechanism in SNR shocks only if the Alfv\'{e}nic Mach number is smaller than $\sim 20$. This result implies that significant magnetic field amplification must occur in the upstream region of SNR shocks. Indeed, if the upstream field is not amplified, the typical Alfv\'{e}nic Mach number for SNR shocks would be $\sim 300$, for a shock with $v_{sh}=3000$ km/s propagating in a $n_i=1$ cm$^{-3}$ plasma with a $3 \mu$G ISM field. This is interesting considering that significant magnetic amplification in SNR shocks based on X-ray observations has been recently reported \citep{Ballet06, UchiyamaEtAl07}. These observations show the existence of thin, non-thermal rims, which are interpreted as synchrotron emission by TeV electrons accelerated at the shocks. The rapid variability and thinness of the rims (which depend on the synchrotron cooling time of the electrons) have allowed to estimate the strength of the field, suggesting downstream amplitudes $\sim 100$ times larger than typically expected in the ISM of the Galaxy. This implies an amplification factor of the upstream field of $\sim 25$, considering that the field compression at the shock itself would contribute an extra factor of $\sim 4$ to the total downstream field growth, making it possible to have $M_A \lesssim 20$ in these shocks. Also, there are theoretical reasons to believe that this field growth could happen in the upstream medium of the shocks, due to streaming instabilities driven by cosmic rays (CRs) as they get accelerated in these environments \citep{Bell04, Bell05, RiquelmeEtAl09, RiquelmeEtAl10}.

The $M_A \lesssim 20$ condition for electron injection is also supported by a recent PIC study of a two-dimensional, perpendicular shock with in-plane magnetic field and parameters: $M_A \approx 130$, $v_{in}=0.25c$, and $m_i/m_e=30$ \citep{KatoEtAl10}. This study showed the absence of electron acceleration and whistler waves in the foot of the shock. Indeed, the shock foot is dominated by the Weibel instability, which plays a fundamental role in the formation of the shock itself. Although the strict perpendicularity of the shock ($\theta_{Bn}=90^{\circ}$) may contribute to reducing the electron acceleration, we believe that the suppression of the whistler waves in this low magnetization shock would make electron injection unlikely for any value of $\theta_{Bn}$.

Thus, this result reinforces the idea of very large magnetic amplification associated with the synchrotron rims seen by the X-ray observations of SNRs. It is important to emphasize that in this paper we have concentrated only on the injection part of the acceleration process. After the injected electrons reach energies such that they can move diffusively in the shock vicinity, they would be further accelerated via the DSA up until the $\sim 10-100$ TeV energies inferred from observations. The DSA part of the acceleration process is not captured in our simulations due to the lack of upstream turbulence.

As seen from Figure \ref{fig:spectconclu}, ions are not accelerated in the quasi-perpendicular configurations studied in this work. On the other hand, in quasi-parallel shocks, Alfv\'{e}n waves would interact resonantly with returning ions, providing the pitch angle scattering necessary to confine them to the shock vicinity, and allowing their acceleration via the first order Fermi process \citep[][Gargate \& Spitkovsky 2010, in prep.]{KulsrudEtAl69}. Thus, given that the acceleration of electrons and ions would require different magnetic orientations, in principle, the acceleration of these two species would not be possible in the same regions of the shocks. This fact poses a consistency problem when applying the electron injection mechanism presented here to the case of SNR shocks, given that in SNR conditions the acceleration of electron requires the presence of relativistic ions to amplify the field. This problem would be solved if the magnetic growth could also involve magnetic field reorientation. This is actually what has been found by numerical studies of the CR streaming instabilities, which are characterized by very turbulent magnetic configurations in their non-linear state \citep{Bell04, Bell05, RiquelmeEtAl09, RiquelmeEtAl10}. Also, these instabilities would produce magnetic field fluctuations on scales comparable to the Larmor radii of the CRs, $R_{L,cr}$. Therefore, given that  the electron injection mechanism presented here operates on length scales comparable to a few $c/\omega_{p,i}$ (which is much smaller than $R_{L,cr}$), it is plausible that electron acceleration can happen locally, in regions where $\vec{B}$ is quasi-perpendicular to the shock normal.

The likely global picture of the shock acceleration process then unfolds as follows. On large scales upstream of the efficiently accelerating shock, the magnetic field  is likely quasi-parallel to the shock normal. This configuration is generally conducive to ion (cosmic ray) acceleration and subsequent escape. Escaping cosmic rays amplify and reorient the upstream magnetic field closer to the shock via current-driven instabilities. The resulting magnetic turbulence advected to the shock will be roughly isotropic, with regions of quasi-perpendicular magnetic field intermittently crossing the shock. This locally transverse field will be efficient in injection of electrons via the whistler mechanism. Once pre-accelerated, the electrons will join the ions in DSA on scales larger than the shock foot in the turbulence that is driven by the cosmic rays. This speculative picture underscores the interdependence of electron and ion acceleration, as without cosmic rays the amplification and reorientation of magnetic field needed for electron injection would be hard to achieve. Observationally, this scenario is consistent with the large scale magnetic field in SN1006 pointing along the axis that connects the ``polar caps" of bright non-thermal emission (Ballet 2006). Also, the degree of linear polarization should be smaller in SNR rims with strong synchrotron emission, as we expect the field direction near the shock to be randomized by the amplified turbulence \citep{Stroman09}. On larger scales, the radial magnetic fields inferred from radio polarization measurements of synchrotron-emitting regions of SNR shocks are also consistent with our model (Dickel et al.~1991, DeLaney et al.~2002).  

The results presented in this paper correspond to only a partial exploration of the space of shock parameters. Studying this problem using realistic mass ratios is computationally expensive, so testing all the possible combinations of parameters of interest is complicated and will require further work. For instance, one of our basic assumptions is that the angle $\theta_{Bn}$ at which the acceleration is maximized has a weak dependence on the ion to electron mass ratio. We believe that this assumption needs to be investigated by performing a more complete exploration of the acceleration efficiency dependence on $\theta_{Bn}$ and $m_i/m_e$. In a similar way, the dependence of the acceleration efficiency on the shock velocity $v_{sh}$ requires further study. We found that, when passing from $v_{sh}=0.14$ to 0.42, the slope of the power law part of the electron spectra for the mass ratios $m_i/m_e=25$, 100, and 400 does not change significantly. However, this result was obtained only in the case of $\theta_{Bn}=75^o$ and $M_A=7$. We think that it would be interesting to confirm this weak velocity dependence by performing a thorough exploration of the shock parameter space at lower velocities. Finally, comparing two-dimensional simulations with analogous three-dimensional runs where the typical whistler length scale ($\sim 10c/\omega_{p,e}$) is resolved in the third dimension would also be useful to confirm our results and to enquire about possible additional three-dimensional effects. Although we do not expect the qualitative picture presented here to change significantly, we believe that these studies would give us more accuracy in our estimates of the electron injection efficiencies as a function of the shock parameters. 

With these caveats in mind, in this paper we have identified a new possible mechanism for electron injection into the DSA process in non-relativistic shocks. The obtained spectra are consistent with in-situ measurements of electron energy distributions at the Earth's bow shock. Also, our results would explain the observed electron acceleration in SNR shocks, implying very large upstream magnetic field amplification in these environments. Thus, if this mechanism proves to be the only possible solution for electron injection in non-relativistic blast waves, it would reinforce the inferred strong connection between particle acceleration and magnetic field amplification in SNR shocks.   

\acknowledgements

M.~A.~R. thanks the support from the Department of Astrophysical Sciences of Princeton University. This research was supported by NSF grant AST-0807381. The simulations presented in this article were performed on computational resources supported by the PICSciE-OIT High Performance
Computing Center and Visualization Laboratory of Princeton University. This research also used resources of the National Energy Research Scientific Computing Center, which is supported by the Office of Science of the U.S. Department of Energy under Contract No. DE-AC02-05CH11231.
  

\appendix

\section{Mass Ratio Dependence of the ``Shock Surfing" Acceleration}
\label{sec:appendix}
\begin{figure*}[t!]
\centering
\includegraphics[width=8cm]{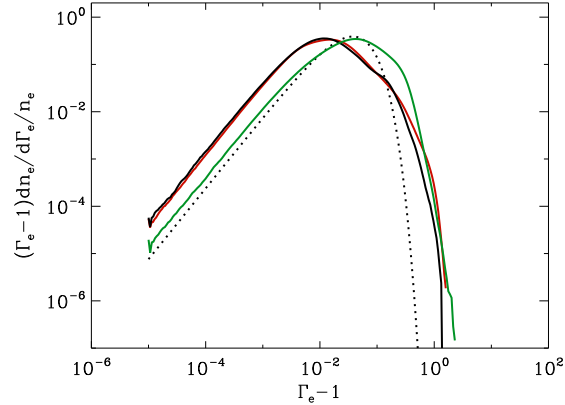}
\caption{The downstream electron spectra at $t\omega_{c,i}=10$ for the two-dimensional simulations 2Db-1 (red line), 2Db-2 (black line), and 2Db-3 (green line), described in Table \ref{table:2Dperp}. The simulations are characterized by having a magnetic field quasi-perpendicular to the simulation plane, and only differ in their mass ratios, which have values  $m_i/m_e=25$, 100, and 400, respectively.
}
\label{fig:compmasspect}
\end{figure*}
\begin{figure}[t!]
\centering
\includegraphics[width=12cm]{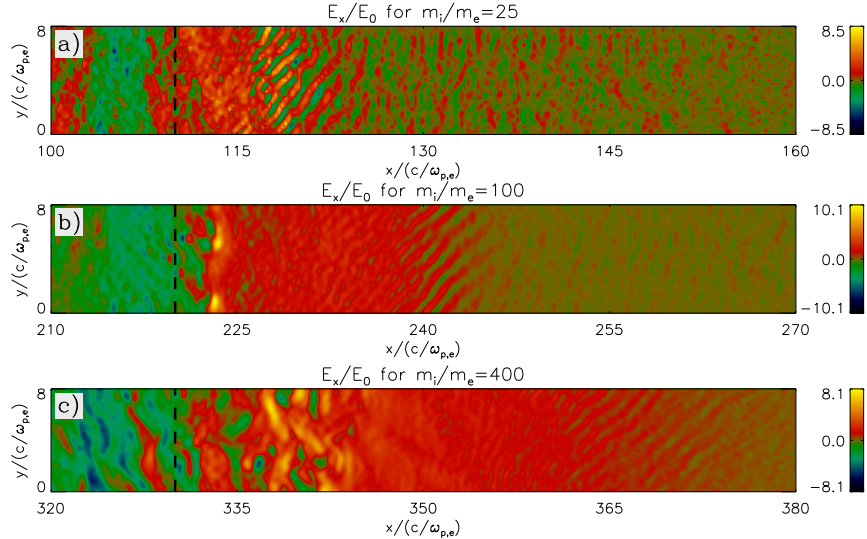}
\caption{The electric field along $\hat{x}$, $E_x$, in the shock transition region of simulations 2Db-1 (panel $a$), 2Db-2 (panel $b$), and 2Db-3 (panel $c$), which are described in Table \ref{table:2Dperp}. These simulations are characterized by having a magnetic field quasi-perpendicular to the simulation plane, and only differ in their $m_i/m_e$ parameter given by $m_i/m_e=$25, 100, and 400, respectively. The field is shown at $t\omega_{c,i}=10$, and is normalized in terms of $E_0$, which corresponds to the convective electric field given by $B_0v_{in}/c$ (where $v_i=0.1c$ is the injection velocity of the upstream plasma as seen from the downstream frame).} 
\label{fig:massbuneman}
\end{figure}
\begin{deluxetable*}{ccccccccc}
\tabletypesize{\scriptsize}
\tablecaption{Parameters for the two-dimensional simulations with magnetic field quasi-perpendicular to simulation plane}
\tablehead{ \colhead{Run} &  \colhead{$c/\omega_{p,e}$}&  \colhead{$L_y/c/\omega_{p,e}$}  &  \colhead{$\beta_e=\beta_i$}&\colhead{$v_{sh}/c$} &
 \colhead{$M_A$}  &  \colhead{$\theta_{Bn}$} &  \colhead{$m_i/m_e$} &  \colhead{N$_{ppc}$}
}
\startdata
2D-1b &  15&  $26$&  0.5& 0.14 & 7 & $75^{\circ}$ & 25& 100\\ 
2D-2b &  15&  $34$&  0.5& 0.14 & 7 & $75^{\circ}$ & 100& 100\\
2D-3b &  10&  $51$&  0.5& 0.14 & 7 & $75^{\circ}$ & 400& 100\\
2D-4b &  10&  $51$&  4.5& 0.14 & 21 & $75^{\circ}$ & 25& 100\\ 
2D-5b &  10&  $26$&  4.5& 0.14 & 21 & $75^{\circ}$ & 100 & 100\\
2D-6b &  15&  $34$&  0.045& 0.042 & 7 & $75^{\circ}$ & 25& 100\\
2D-7b &  10&  $26$&  0.045& 0.042 & 7 & $75^{\circ}$ & 100& $\textrm{ }100$ 
\enddata
\tablecomments{We list the electron skin depth $c/\omega_{p,e}$ in terms of number of grid cells, the transverse size of the simulation box in terms of $c/\omega_{p,e}$, the beta parameter of the different plasma particles $\beta_j$ ($\equiv p_j/B_0^2/8\pi$, where $p_j$ is the pressure of particle ``$j$"), the upstream medium shock velocity, $v_{sh}$, the Alfv\'{e}nic Mach number, $M_A$, the angle between the upstream magnetic field and the shock normal, $\theta_{Bn}$, the ion to electron mass ratio, $m_i/m_e$, and the total number of particles per cell, $N_{ppc}$, in the simulation.}
\label{table:2Dperp}
\end{deluxetable*}


In this appendix we explore the mass ratio dependence of the electron energization due to shock surfing acceleration \citep{AmanoEtAl09}. This acceleration is driven by the presence of electrostatic waves, produced by the Buneman instability \citep{Buneman58}, in the foot of quasi-perpendicular shocks. The Buneman waves grow in the leading edge of the foot due to the relative velocity between the upstream electrons and the shock-reflected ions, and have a typical scale comparable to $\sim c/\omega_{p,e}$. The small length scale of these waves, which can be comparable or even smaller than the electron Larmor radii, makes the Buneman waves a potentially important means for electron scattering and energization in the foot of quasi-perpendicular shocks. As the upstream electrons encounter these waves, they gain energy due to the work performed by the fluctuating electric field, combined with the convective field of the upstream plasma. The wave vector of the fastest growing mode is parallel to the velocity of the beam of returning ions, so these waves are best studied either by three-dimensional simulations or by two-dimensional runs where the magnetic field is quasi-perpendicular to the simulation plane, so that the gyrational motion of the ions is resolved on the plane\footnote{As in the main part of the paper, in this appendix the two-dimensional simulations will be in the $xy$ plane.}

Using a two-dimensional simulation with shock parameters $v_{in}=0.2c$ ($v_{sh}\approx0.3c$), $M_A=14$, $\theta_{Bn}=90^o$ (with $\vec{B}_0$ perpendicular to the simulation plane), and $m_i/m_e=25$, \cite{AmanoEtAl09} showed that an electron spectrum with index $\alpha \approx$ 2-2.5 can be produced in the shock transition region due to this mechanism.

In this appendix we show, however, that the importance of this shock surfing acceleration decreases as the mass ratio increases to more realistic values. As an example, let us compare the downstream spectra of three simulations with the same shock parameters $v_{in}=0.1c$ ($v_{sh}\approx0.14c$), $M_A=7$, $\theta_{Bn}=75^o$ (and the angle between $\vec{B}_0$ and $\hat{z}$ equal to $15^{\circ}$), but different mass ratios: $m_i/m_e=25, 100$, and 400 (the rest of the parameters are specified in Table \ref{table:2Dperp}). Figure \ref{fig:compmasspect} shows the spectra at $t\omega_{c,i}=10$ measured in the downstream medium. Although in all three cases a non-thermal component can be seen, the maximum energy reached by the electrons is about the same ($\Gamma_e-1 \approx 1$), despite the factor of 16 difference in the used mass ratios. In particular, in the $m_i/m_e=400$ case, the closeness between the peak of the thermal part of the distribution and the maximum energy of the accelerated particles reduces significantly the relative fraction of non-thermal particles compared with $m_i/m_e=25$ and 100 cases. Also, notice that in these two-dimensional simulations the acceleration due to whistler waves gets suppressed. This is because these simulations only allow the wave vector of the whistlers to be quasi-perpendicular to the magnetic field, inhibiting their growth and their effect on the electron energization.

In terms of the evolution of the Buneman waves, we also observe a decrease in their relative amplitude. This can be seen from Figure \ref{fig:massbuneman}, where the electric field along the $x-$axis is depicted for the same simulations shown in Figure \ref{fig:compmasspect}. The Buneman waves correspond to the fluctuations on $\sim c/\omega_{p,e}$ scale that appear in the leading edge of the shock (at $x/c/\omega_{p,e}=115-120$, 240-245, and 365-370, in panels $a)$, $b)$, and $c)$, respectively). The black dashed line in the three panels marks the position of the shock density peak (overshoot) in the three simulations. When compared to the convective electric field ($E_0=B_0v_{in}/c$), we see that the maximum electric field of the Buneman waves is reduced by a factor of $\sim 4$ when $m_i/m_e$ passes from 25 to 400.

The same maximum energy for the accelerated electrons is observed in similar simulations but using $M_A=21$ (simulations 2Db-4 and 2Db-5). Also, when trying cases similar to the ones shown in Figures \ref{fig:compmasspect} and \ref{fig:massbuneman} but using $v_{sh}=0.042$ (simulations 2Db-6 and 2Db-7), a decrease in the maximum energy to $\Gamma_e-1\approx 0.1$ is seen, with no change by varying $m_i/m_e$.

Thus, in general, the maximum energy attainable by the shock surfing mechanism appears to depend mainly on the shock velocity, and shows no changes for different values of $m_i/m_e$. Therefore, as the mass ratio $m_i/m_e$ is increased, the peak of the thermal part of the distribution gets closer to this maximum energy. This implies that the fraction of non-thermal electrons is reduced when $m_i/m_e$ approaches more realistic values. We conclude that this mechanism would not contribute significantly to electron acceleration in the fully realistic case of $m_i/m_e=1836$. 

\end{document}